\RequirePackage{fix-cm}
\documentclass[smallextended]{svjour3}       

\smartqed  

\usepackage{graphicx}

\usepackage{multirow}
\usepackage{epsfig}
\usepackage{enumerate}

\journalname{Foundations of Physics, {\bf 41}, doi:10.1007/s10701-011-9534-7 (2011)}

\title{Parity proofs of the Bell-Kochen-Specker theorem based on the 600-cell}

\author{Mordecai Waegell \and P.K. Aravind \and Norman D. Megill
\and Mladen Pavi\v ci\'c}

\authorrunning{M. Waegell \and P.K. Aravind \and
N.D. Megill \and M. Pavi\v ci\'c}

\institute{Mordecai Waegell \and P.K. Aravind \at
Physics Department, Worcester Polytechnic Institute,
Worcester, MA 01609, U.S.A.\\
\email{caiw@wpi.edu} $\cdot$ \email{paravind@wpi.edu}
\and Norman D.Megill \at
Boston Information Group, 19 Locke Ln. Lexington, MA 02420, U.S.A.
\email{nm@alum.mit.edu}
\and Mladen Pavi\v ci\'c \at
 Institute for Theoretical Atomic, Molecular
and Optical Physics at Physics Department at
Harvard University and Harvard-Smithsonian Center for
Astrophysics, Cambridge, MA 02138, USA \&\
Chair of Physics, Faculty of Civil Engineering, University
of Zagreb,  Croatia.\\
\email{mpavicic@grad.hr}}

\date{\today}

\begin{document}
\maketitle
\begin{abstract}
      The set of 60 real rays in four dimensions
derived from the vertices of a 600-cell is shown
to possess numerous subsets of rays and bases that
provide basis-critical parity proofs of the Bell-Kochen-Specker
(BKS) theorem (a basis-critical proof is one that fails
if even a single basis is deleted from it).
The proofs vary considerably in size, with the smallest having 26
rays and 13 bases and the largest 60 rays
and 41 bases. There are at least 90 basic types
of proofs, with each coming in a number
of geometrically distinct varieties. The replicas of all the
proofs under the symmetries of the 600-cell yield
a total of almost a hundred million parity proofs
of the BKS theorem. The proofs are all very transparent
and take no more than simple counting to verify.
A few of the proofs are exhibited, both in tabular form
as well as in the form of MMP hypergraphs that
assist in their visualization. A survey of the proofs is
given, simple procedures for generating some of
them are described and their applications are discussed.
It is shown that all four-dimensional parity proofs
of the BKS theorem can be turned into experimental
disproofs of noncontextuality.
\end{abstract}

\section{\label{sec:intro}Introduction}

           In a recent paper \cite{Waegell2010}
two of us showed that the system of 60 rays derived
from the vertices of a 600-cell could be used to
give two new proofs of the Bell-Kochen-Specker (BKS)
theorem \cite{Bell1966,KS1967} ruling out the existence
of noncontextual hidden variables theories. A later
work \cite{Pavicic2010} presented several additional
proofs based on the same set of rays. The purpose
of this paper is to add to the store of proofs in
\cite{Pavicic2010}, but, even more than that, to
convey a feeling for the variety and flavor of the
proofs (both through the examples presented here
and the far more extensive listing on the website
\cite{WebsitePKA2010}) and to show interested readers
how many of the proofs can be obtained by simple
constructions based on the geometry of the 600-cell.
There are two aspects of the present proofs that
make them noteworthy. The first is that they are
all ``parity proofs'' (this term is explained in
Sec.~\ref{sec:parity-proof}) whose validity can
be checked by simple counting.  And the second is
that there are about a hundred million of them in
this 60-ray system. While many of the proofs are
just replicas of each other under the symmetries
of the 600-cell, the number of distinct proofs,
in terms of size and other characteristics,
is still fairly large (we used a random exhaustive
generation of proofs to obtain over 8,000 proofs,
most of which turned out to be parity proofs ~\cite{Aravind2010}).
The sheer profusion and variety of parity proofs
contained in the 600-cell is unmatched by that in
any other system we are aware of and motivated us
to study this system in detail, both for its geometric
interest as well as for its possible applications.

        A brief survey of earlier proofs of the
BKS theorem may be helpful in setting the present
work in context. After Kochen and Specker \cite{KS1967}
first gave a finitary proof of their theorem using
117 directions in ordinary three-dimensional space,
a number of authors gave alternative proofs in three
\cite{Peres1991,PeresBook,Bub,Penrose2000}, four
 \cite{Peres1991,Cabello1996,Kernaghan1994,ZP1993,pmmm03a,pmmm04b,pmm-2-09}
and higher \cite{Kernaghan1995,CKReview,DivPeres,Ruuge}
dimensions. Some of the proofs in higher dimensions
are much simpler than the three-dimensional proofs
and, in fact, are examples of the ``parity proofs''
we discuss throughout this paper. In recent years
there has been a resurgence of interest in the BKS
theorem as a result of the fruitful suggestion by
Cabello \cite{Cabello2008} of how it might be experimentally
tested. Cabello's basic observation is that many
proofs of the BKS theorem based on a finite set of
rays and bases can be converted into an inequality
that must be obeyed by a noncontextual hidden variables
theory but is violated by quantum mechanics. Experimental
tests of Cabello-like inequalities have been carried
out in four-level systems realized by ions \cite{Kirchmair},
neutrons \cite{Bartosik}, photons \cite{Amselem}
and nuclear spins \cite{Moussa}, and violations
of the inequalities have been observed in all the
cases. Still other inequalities, some state-dependent and
others not, that must be satisfied by noncontextual theories have been derived for
qutrits \cite{Klyachko}, n-qubit systems \cite{Cabello2010b} and hypergraph models
\cite{Severini}. It has been argued in \cite{Cabello2010a} that contextuality
is the key feature underlying quantum nonlocality. A wide ranging discussion of the
Kochen-Specker and other no-go theorems, as well as the subtle interplay between the
notions of contextuality, nonlocality and complementarity, can be found in \cite{Liang}.
Aside from their foundational interest, proofs of the BKS theorem are useful in
connection with protocols such as quantum cryptography
\cite{BPPeres}, random number generation \cite{Svozil}
and parity oblivious transfer \cite{Spekkens}.

        This paper is organized as follows. Sec.~\ref{sec:parity-proof}
reviews the BKS theorem and explains what is meant
by a ``parity proof'' of it. An explanation is also
given of the notion of a ``basis-critical'' parity
proof, since only such proofs are presented in this
paper. Sec.~\ref{sec:parity-overview} introduces
the system of 60 rays and 75 bases derived from
the 600-cell that is the source of all the proofs
presented in this paper. A notation is introduced
for the ray-basis sets underlying the parity proofs,
and an overview is given of all the parity proofs
we were able to find in the 600-cell. The algorithm
we used to search for the proofs is described, and
a few of the proofs are displayed in a tabular form
so that the reader can see how they work. An equivalent
McKay-Megill-Pavicic (MMP) hypergraph representation
\cite{Pavicic2010,pmmm03a,bdm-ndm-mp-fresl-jmp-10}
is used to give the reader a graphical visualisation
of some of the proofs. In an MMP hypergraph vertices
correspond to rays and edges to tetrads of
mutually orthogonal rays (see
Figs.~\ref{fig:30-15-a-b} and \ref{fig:3417-2613}).
Sec.~\ref{sec:discussion} summarizes the general
features of the parity proofs and also points out
their relevance for quantum key distribution and
experimental disproofs of noncontextuality. The
Appendix reviews some basic geometrical facts about
the 600-cell and shows how they can be used to give
simple constructions for some of the parity proofs
in Table \ref{Overview}. Space prevents us from
discussing more than a handful of examples,
but the ones chosen may help to convey some feeling
for the rest. Some virtues of the treatment in the
Appendix are (a) that it allows many of the proofs
to be constructed ``by hand'' without the need to
look up a compilation, (b) that it allows the number
of replicas of a particular proof under the symmetries
of the 600-cell to be determined, and (c) that it
reveals close connections between different proofs
that might otherwise appear to be unrelated. However
the treatment in the Appendix is not needed for
an understanding of the main results of this paper
and can be omitted by those not interested in it.

        This paper is written to be self-contained
and can be read without any knowledge of our earlier
work \cite{Waegell2010,Pavicic2010} on this problem.

\section{\label{sec:parity-proof}Parity proofs of the BKS theorem; basis critical sets}

The BKS theorem asserts that in any Hilbert space
of dimension $d\geq 3$ it is always possible to
find a finite set of rays \cite{note1} that cannot
each be assigned the value $0$ or $1$ in such a
way that (i) no two orthogonal rays are both assigned
the value 1, and (ii) not all members of a basis,
i.e. a set of $d$ mutually orthogonal rays, are
assigned the value 0. The proof of the theorem becomes
trivial if one can find a set of $R$ rays in $d$
dimensions that form an odd number, $B$, of bases
in such a way that each ray occurs an even number
of times among those bases. Then the assignment
of 0's and 1's to the rays in accordance with rules
(i) and (ii) is seen to be impossible because the
total number of 1's over all the bases is required
to be both odd (because each basis must have exactly
one ray labeled 1 in it) and even (because each
ray labeled 1 is repeated an even number of times).
Any set of $R$ rays and $B$ bases that gives this
even-odd contradiction furnishes what we call a
``parity proof'' of the BKS theorem.

          Let us denote a set of $R$ rays that forms
$B$ bases a $R$-$B$ set. A $R$-$B$ set that yields
a parity proof of the BKS theorem will be said to
be basis-critical (or simply critical) if dropping
even a single basis from it causes the BKS proof
to fail. Basis-criticality is not to be confused
with ray-criticality, which takes all orthogonalities
between rays into account and not just those in
the limited set of bases considered. We focus on
basis-criticality because it is more relevant to
experimental tests of the Kochen-Specker theorem.
Such tests typically involve projective measurements
that pick out whole sets of bases, and performing
a test that corresponds to a basis-critical set
is an efficient strategy because it involves no
superfluous measurements. The only parity proofs
exhibited in this paper are those that correspond
to basis-critical sets.

\section{\label{sec:parity-overview}Overview
of parity proofs contained in the 600-cell}

       Table \ref{tab:Ray} shows the 60 rays derived
from the vertices of the 600-cell and
Table \ref{tab:Basis} the 75 bases (of four rays
each) formed by them. Each ray occurs in exactly
five bases, with its 15 companions in these bases being
the only other rays it is orthogonal to. Thus
Table \ref{tab:Basis}  (or the ``basis table'')
captures all the orthogonalities between the
rays and is completely equivalent to their Kochen-Specker
diagram.

\begin{table}[htp]
\vskip-12pt
\addtolength{\tabcolsep}{10pt}
\centering 
\begin{tabular}{|c c c c|} 
\hline 
1 = $2 0 0 0 $ & 2 = $0 2 0 0 $ & 3 = $0 0 2 0 $ & 4 = $0 0 0 2 $ \\
5 = $1 1 1 1 $ & 6 = $1 1 \overline{1} \overline{1} $ &
7 = $1 \overline{1} 1 \overline{1} $ & 8 = $1 \overline{1} \overline{1} 1 $ \\
9 = $1 \overline{1} \overline{1} \overline{1} $ &
10 = $1 \overline{1} 1 1 $ & 11 = $1 1 \overline{1} 1 $ &
12 = $1 1 1 \overline{1} $ \\
\hline
13 = $\kappa 0 \overline{\tau} \overline{1} $ &
14 = $0 \kappa 1 \overline{\tau} $ &
15 = $\tau \overline{1} \kappa 0 $ & 16 = $1 \tau 0 \kappa $ \\
17 = $\tau \kappa 0 \overline{1} $ & 18 = $1 0 \kappa \tau $ &
19 = $\kappa \overline{\tau} \overline{1} 0 $ &
20 = $0 1 \overline{\tau} \kappa $ \\
21 = $1 \kappa \tau 0 $ & 22 = $\tau 0 \overline{1} \kappa $ &
23 = $0 \tau \overline{\kappa} \overline{1} $ &
24 = $\kappa \overline{1} 0 \overline{\tau} $ \\
\hline
25 = $\tau 0 1 \kappa $ & 26 = $0 \tau \overline{\kappa} 1 $ &
27 = $1 \overline{\kappa} \overline{\tau} 0 $ &
28 = $\kappa 1 0 \overline{\tau} $ \\
29 = $0 \kappa 1 \tau $ & 30 = $\tau 1 \overline{\kappa} 0 $ &
31 = $\kappa 0 \tau \overline{1} $ & 32 = $1 \overline{\tau} 0 \kappa $ \\
33 = $\tau \overline{\kappa} 0 \overline{1} $ &
34 = $0 1 \overline{\tau} \overline{\kappa} $ &
35 = $1 0 \overline{\kappa} \tau $ & 36 = $\kappa \tau 1 0 $ \\
\hline
37 = $\tau 0 \overline{1} \overline{\kappa} $ &
38 = $0 \tau \kappa \overline{1} $ & 39 = $1 \overline{\kappa} \tau 0 $ &
40 = $\kappa 1 0 \tau $ \\
41 = $\tau 1 \kappa 0 $ & 42 = $0 \kappa \overline{1} \overline{\tau} $ &
43 = $1 \overline{\tau} 0 \overline{\kappa} $ &
44 = $\kappa 0 \overline{\tau} 1 $ \\
45 = $0 1 \tau \kappa $ & 46 = $\tau \overline{\kappa} 0 1 $ &
47 = $\kappa \tau \overline{1} 0 $ & 48 = $1 0 \kappa \overline{\tau} $ \\
\hline
49 = $\kappa 0 \tau 1 $ & 50 = $0 \kappa \overline{1} \tau $ &
51 = $\tau \overline{1} \overline{\kappa} 0 $ &
52 = $1 \tau 0 \overline{\kappa} $ \\
53 = $1 0 \overline{\kappa} \overline{\tau} $ &
54 = $\tau \kappa 0 1 $ & 55 = $0 1 \tau \overline{\kappa} $ &
56 = $\kappa \overline{\tau} 1 0 $ \\
57 = $\tau 0 1 \overline{\kappa} $ &
58 = $1 \kappa \overline{\tau} 0 $ &
59 = $\kappa \overline{1} 0 \tau $ & 60 = $0 \tau \kappa 1 $ \\
\hline 
\end{tabular}
\caption{The 60 rays of the 600-cell. The numbers
following each ray are its components in an orthonormal
basis, with $\tau=(1+\surd5)/2$, $\kappa=1/\tau$,
a bar over a number indicating its negative and
commas being omitted between components. The entries can
also be regarded as coordinates of 60 of the vertices
of a 600-cell, located on a sphere of radius 2 centered
at the origin. The other 60 vertices are the antipodes of these.}
\vskip-15pt
\label{tab:Ray} 
\end{table}

\begin{table}[htp]
\addtolength{\tabcolsep}{-2.3pt}
\begin{center}
\begin{tabular}{|c|cccc|cccc|cccc|cccc|cccc|}
\hline
&\multicolumn{4}{|c|}{A}&\multicolumn{4}{|c|}{B}&\multicolumn{4}{|c|}{C}&
\multicolumn{4}{|c|}{D}&\multicolumn{4}{|c|}{E}\\
\hline
\multirow{3}{*}{A\'}&1&2&3&4&31&42&51&16&22&60&39&28&57&23&27&40&44&29&15&52\\[-1.5pt]
&5&6&7&8&38&24&58&25&18&47&33&55&36&53&20&46&59&26&37&21\\[-1.5pt]
&9&10&11&12&56&45&17&35&13&32&50&41&43&49&30&14&34&19&48&54\\
\hline
\multirow{3}{*}{B\'}&13&14&15&16&43&54&3&28&34&12&51&40&9&35&39&52&56&41&27&4\\[-1.5pt]
&17&18&19&20&50&36&10&37&30&59&45&7&48&5&32&58&11&38&49&33\\[-1.5pt]
&21&22&23&24&8&57&29&47&25&44&2&53&55&1&42&26&46&31&60&6\\
\hline
\multirow{3}{*}{C\'}&25&26&27&28&55&6&15&40&46&24&3&52&21&47&51&4&8&53&39&16\\[-1.5pt]
&29&30&31&32&2&48&22&49&42&11&57&19&60&17&44&10&23&50&1&45\\[-1.5pt]
&33&34&35&36&20&9&41&59&37&56&14&5&7&13&54&38&58&43&12&18\\
\hline
\multirow{3}{*}{D\'}&37&38&39&40&7&18&27&52&58&36&15&4&33&59&3&16&20&5&51&28\\[-1.5pt]
&41&42&43&44&14&60&34&1&54&23&9&31&12&29&56&22&35&2&13&57\\[-1.5pt]
&45&46&47&48&32&21&53&11&49&8&26&17&19&25&6&50&10&55&24&30\\
\hline
\multirow{3}{*}{E\'}&49&50&51&52&19&30&39&4&10&48&27&16&45&11&15&28&32&17&3&40\\[-1.5pt]
&53&54&55&56&26&12&46&13&6&35&21&43&24&41&8&34&47&14&25&9\\[-1.5pt]
&57&58&59&60&44&33&5&23&1&20&38&29&31&37&18&2&22&7&36&42\\
\hline
\end{tabular}
\end{center}
\caption{The 75 bases formed by the 60 rays of the
600-cell; rays are numbered as in Table~\ref{tab:Ray}.}
\vskip-20pt
\label{tab:Basis}
\end{table}

         The rays and bases of the 600-cell make
up a 60-75 set ( i.e., one with 60 rays and 75 bases).
This set does not give a parity proof, but  contains
a large number of subsets that do. A $R$-$B$ subset
of the 60-75 set that yields a parity proof must
have each of its rays occur either twice or four
times among its bases (these being the only possibilities
for the 600-cell). It is easy to see that the number
of rays that occur four times is $2B - R$, while
the number that occur twice is $2R - 2B$.

Table \ref{Overview} gives an overview of all the
parity proofs we have found in the 600-cell. The
smallest proof is provided by a 26-13 set (in which
all 26 rays occur twice each among the bases) and
the largest by a 60-41 set (in which 38 rays occur
twice each and 22 rays four times each among the
bases). Moving one step to the left in any row of
Table \ref{Overview} causes the number of rays that
occur four times to go up by 1 and the number
that occur twice to go down by 2.

\begin{table}[ht]
\vskip-12pt
\addtolength{\tabcolsep}{3.5pt}
\centering 
\begin{tabular}{|c|ccccccccccc|} 
\hline
B &  \multicolumn{11}{|c|}{R}\\
\hline
 13 & 26 & & & & & & & & & & \\[-1.2pt]
15 & 30 & & & & & & & & & & \\[-1pt]
17 & 32 & 33 & 34 & & & & & & & & \\[-1pt]
19 & 36 & 37 & 38 & & & & & & & & \\[-1pt]
21 & 38 & 39 & 40 & 41 & 42 & & & & & & \\[-1pt]
23 & 40 & 41 & 42 & 43 & 44 & 45 & 46 & & & & \\[-1pt]
25 & 42 & 43 & 44 & 45 & 46 & 47 & 48 & 49 & 50 & & \\[-1pt]
27 & 44 & 45 & 46 & 47 & 48 & 49 & 50 & 51 & 52 & 53 & 54 \\[-1pt]
29 & 46 & 47 & 48 & 49 & 50 & 51 & 52 & 53 & 54 & 55 & 56\\[-1pt]
31 & 48 & 49 & 50 & 51 & 52 & 53 & 54 & 55 & 56 & 57 & 58 \\[-1pt]
33 & 51 & 52 & 53 & 54 & 55 & 56 & 57 & 58 & 59 & 60 &\\[-1pt]
35 & 53 & 54 & 55 & 56 & 57 & 58 & 59 & 60 & & & \\[-1pt]
37 & 55 & 56 & 57 & 58 & 59 & 60 & & & & & \\[-1pt]
39 & 58 & 59 & 60 & & & & & & & &\\[-1pt]
41 & 60 & & & & & & & & & &\\
\hline
\end{tabular}
\caption{Overview of basis-critical parity proofs
in the 600-cell. Each row shows all the $R$-$B$
parity proofs for a fixed value of $B$ and variable
$R$ ($R$ = number of rays, $B$ = number of bases.)\newline} 
\label{Overview} 
\vskip-30pt
\end{table}

Most of the sets in Table \ref{Overview} were discovered
through a computer search.
The search algorithm is exhaustive, and quite simple:
it starts from an arbitrary basis in Table \ref{tab:Basis} and adds one
basis at a time in an attempt to obtain a target parity proof,
$R$-$B$. Because every ray must appear two or four times in the
proof, a ray appearing once or thrice among the bases already
chosen is selected, and one of the (at most four) other bases
containing that ray is added to the proof at each iteration.  The
algorithm explores all these possible bases-choices in a
branching fashion, and saves computational time by skipping all
branches in which the target $R$ is exceeded before
$B$ bases have been chosen or those in which more than $2B-R$ rays
appear three or four times or those in which any ray appears five
times. This ensures that the search is exhaustive, and keeps the
number of necessary iterations well below the upper bound of $4^B$. If
any branch leads to a $R$-$B$ set, then it is necessarily a parity
proof, while if no branch produces one then the 600-cell contains no
parity proofs of the target size $R$-$B$. The search becomes slower
with increasing values of $R$ and/or $B$, and also as the number of
rays occurring four times in the target set increases, and so we
were not able to carry out the search exhaustively
for all values of $R$ and $B$.
Additional calculations were done after the initial
search to eliminate sets that
corresponded to duplicate or non-critical proofs.

We now give a couple of examples of parity proofs.
A first example, given in Table~\ref{30_15}, 
shows two 30-15 proofs complementary to
each other in the sense that they have no rays in
common. There are exactly 120 such pairs of complementary
proofs, and a simple construction for them is given
in the Appendix. 

\begin{table}[htp]
\vskip-10pt
\addtolength{\tabcolsep}{-3pt}
\renewcommand{\arraystretch}{1.2}
\begin{center}
\begin{tabular}{|c|cccc|cccc|cccc|cccc|cccc|}
\hline
&\multicolumn{4}{|c|}{A}&\multicolumn{4}{|c|}{B}&
\multicolumn{4}{|c|}{C}&\multicolumn{4}{|c|}{D}&\multicolumn{4}{|c|}{E}\\[-1pt]
\hline
\multirow{3}{*}{A'}&1&2&3&4&\phantom{\bf 88}&
\phantom{\bf 88}&\phantom{\bf 88}&\phantom{\bf 88}&\bf 22&\bf 60&\bf
39&\bf 28&\bf 57&\bf 23&\bf 27&\bf 40&\phantom{\bf 88}&\phantom{\bf
88}&\phantom{\bf 88}&\phantom{\bf 88}\\[-3pt]
&&&&&&&&&18&47&33&55&36&53&20&46&&&&\\[-3pt]
&\bf 9&\bf 10&\bf 11&\bf 12&&&&&&&&&&&&&&&&\\[-1pt]
\hline
\multirow{3}{*}{B'}&13&14&15&16&&&&&\bf 34&\bf 12&\bf 51&\bf
40&\bf 9&\bf 35&\bf 39&\bf 52&&&&\\[-3pt]
&&&&&&&&&30&59&45&7&48&5&32&58&&&&\\[-3pt]
&\bf 21&\bf 22&\bf 23&\bf 24&&&&&&&&&&&&&&&&\\[-1pt]
\hline
\multirow{3}{*}{C'}&\bf 25&\bf 26&\bf 27&\bf 28&&&&&&&&&&&&&&&&\\[-3pt]
&29&30&31&32&&&&&\bf 42&\bf 11&\bf 57&\bf 19&\bf 60&\bf
17&\bf 44&\bf 10&&&&\\[-3pt]
&&&&&&&&&37&56&14&5&7&13&54&38&&&&\\[-1pt]
\hline
\multirow{3}{*}{D'}&&&&&&&&&58&36&15&4&33&59&3&16&&&&\\[-3pt]
&\bf 41&\bf 42&\bf 43&\bf 44&&&&&&&&&&&&&&&&\\[-3pt]
&45&46&47&48&&&&&\bf 49&\bf 8&\bf 26&\bf 17&\bf 19&\bf
25&\bf 6&\bf 50&&&&\\[-1pt]
\hline
\multirow{3}{*}{E'}&\bf 49&\bf 50&\bf 51&\bf 52&&&&&&&&&&&&&&&&\\[-3pt]
&53&54&55&56&&&&&\bf 6&\bf 35&\bf 21&\bf 43&\bf 24&\bf
41&\bf 8&\bf 34&&&&\\[-3pt]
&&&&&&&&&1&20&38&29&31&37&18&2&&&&\\[-1pt]
\hline
\end{tabular}
\end{center}
\caption{Two complementary 30-15 parity proofs, one in plain type and the other in bold.}
\vskip-20pt
\label{30_15}
\end{table}

\begin{figure}[htp]
\vskip-15pt
\begin{center}
\includegraphics[height=0.39\textwidth,width=0.4\textwidth]{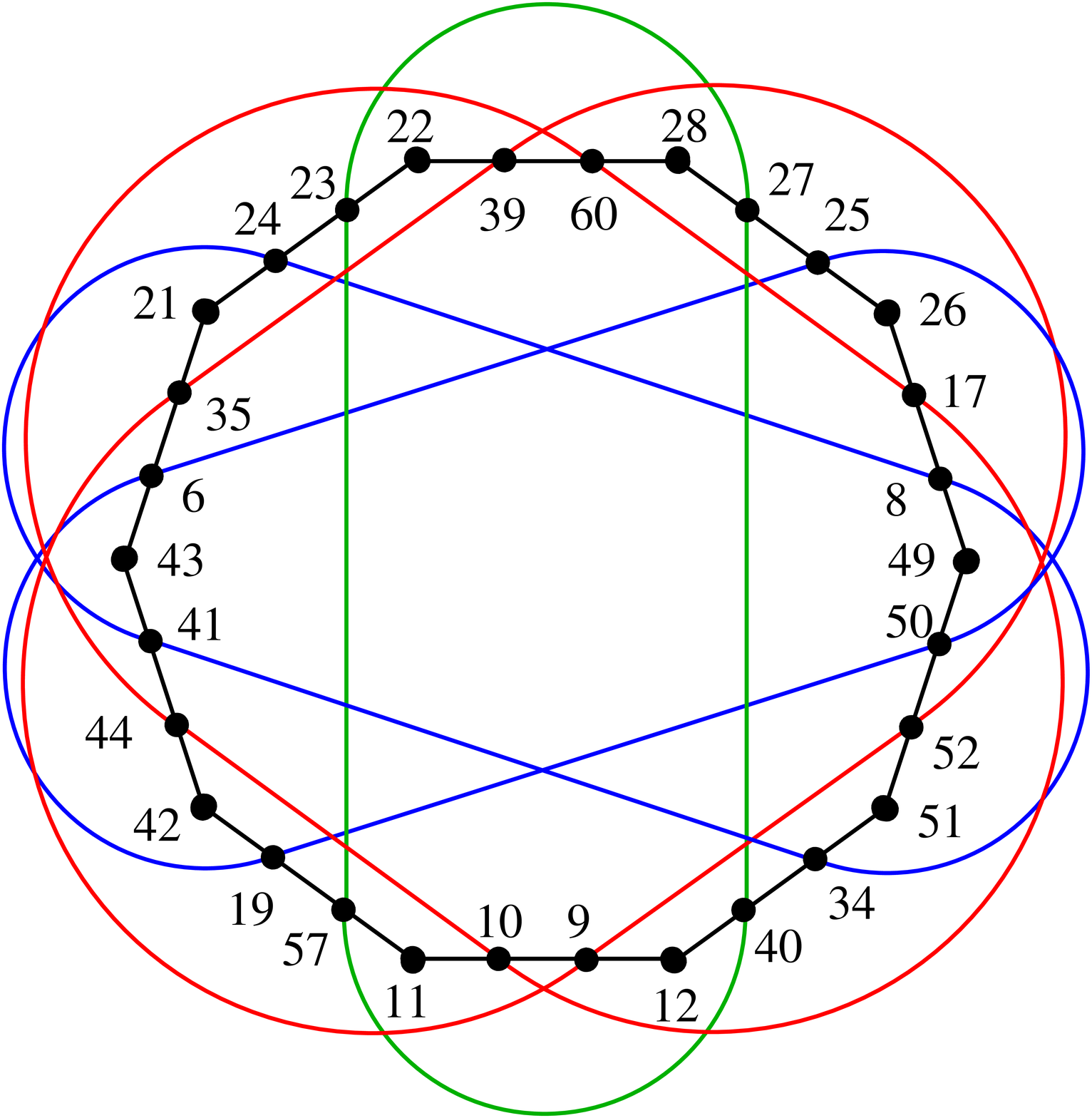}\hbox to 30pt{\hfill}
\includegraphics[height=0.39\textwidth,width=0.4\textwidth]{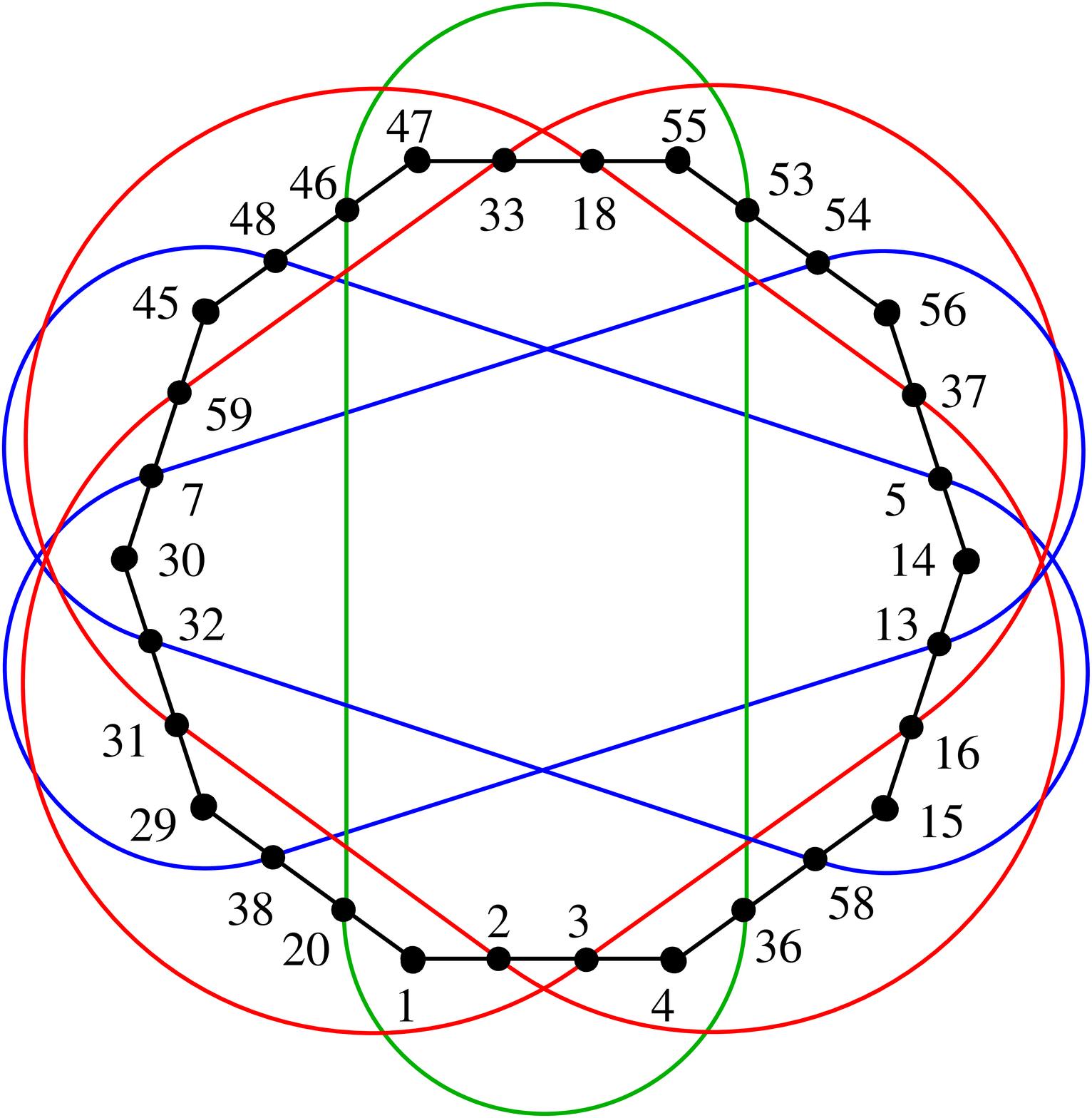}
\end{center}
\caption{MMP hypergraphs~\cite{bdm-ndm-mp-fresl-jmp-10} for the
two 30-15 proofs shown in Table \ref{30_15}. The one on the left
corresponds to the bold bases in Table \ref{30_15}
and the one on the right to the bases in regular
type. Although these two hypergraphs have the same
structure, the 30-15 sets they describe are geometrically
distinct for the reason discussed in Sec. \ref{sec:discussion}(ii).}
\label{fig:30-15-a-b}
\end{figure}

\eject
Figure \ref{fig:30-15-a-b} shows
an alternative representation of the two proofs
in Table \ref{30_15} by means of MMP
hypergraphs.~\cite{bdm-ndm-mp-fresl-jmp-10}
The skeleton for each of the hypergraphs is a decagon
whose alternate sides are bases from a single column
of Table \ref{30_15}. The skeleton is then completed
by five loops crisscrossing the figure that pick
out the bases in the remaining column of Table \ref{30_15}.

\begin{figure}[htp]
\vskip-20pt
\begin{center}
\includegraphics[height=0.4\textwidth,width=0.4\textwidth]{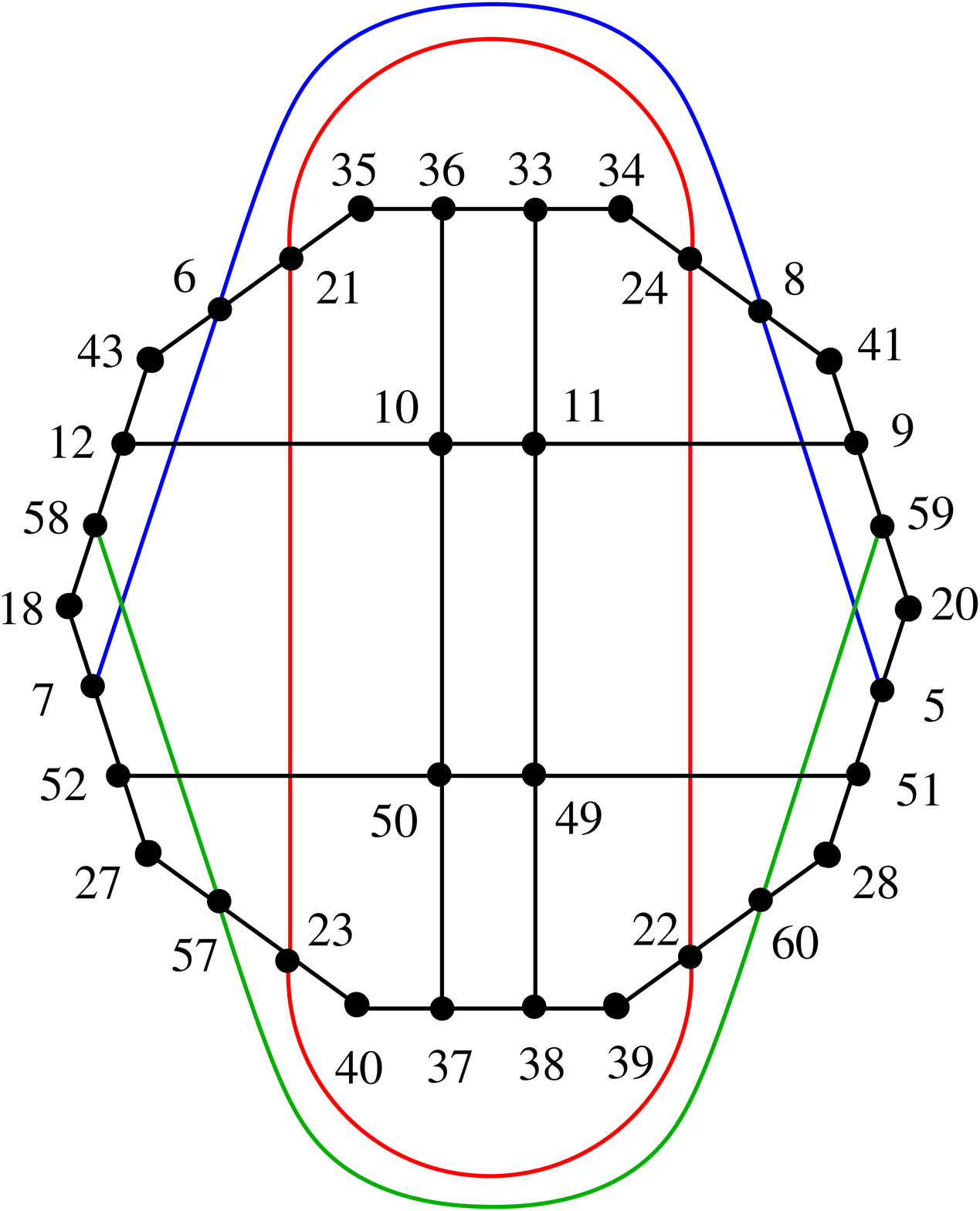}\hbox to 30pt{\hfill}
\includegraphics[width=0.4\textwidth]{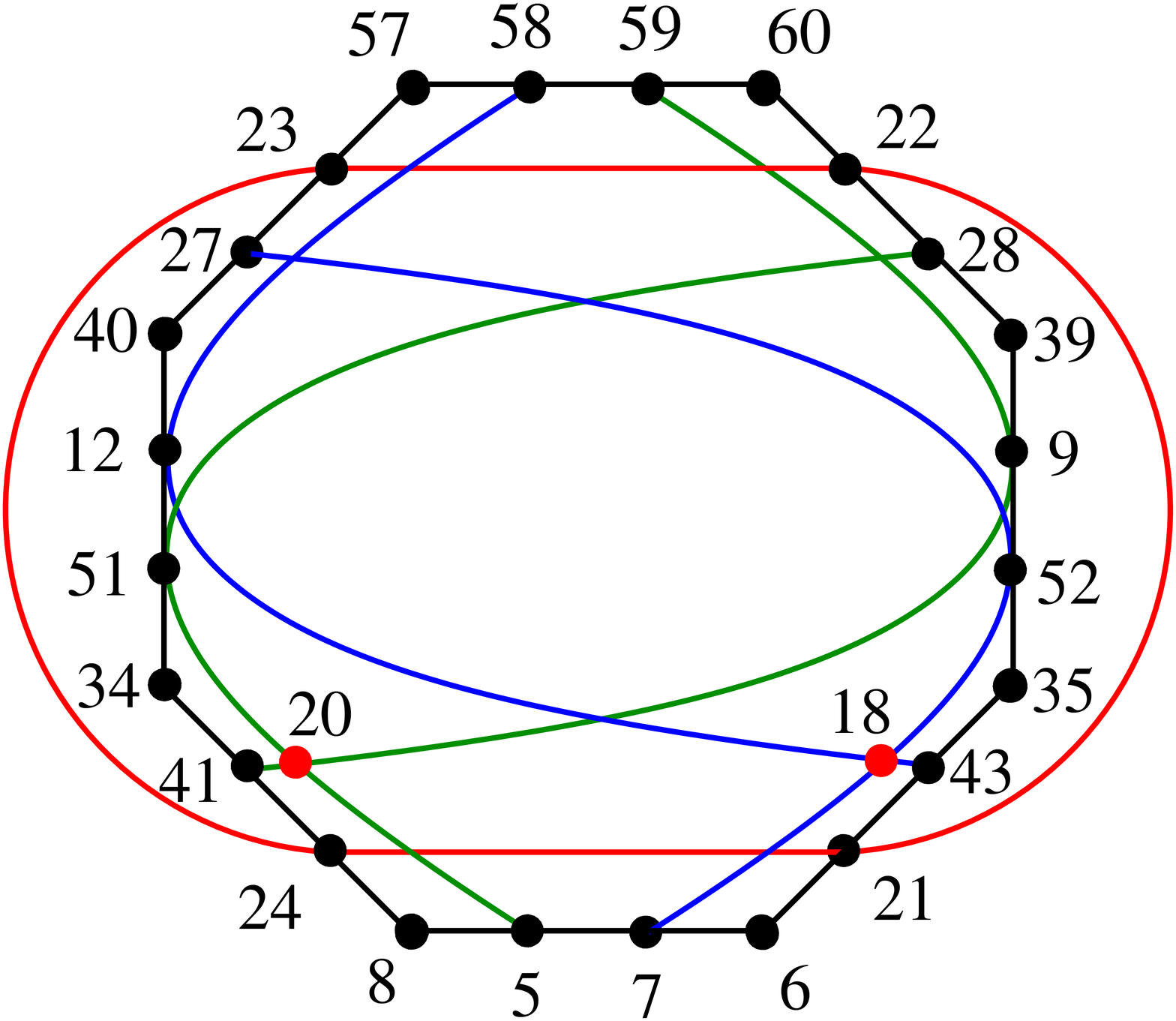}
\end{center}
\caption{MMP hypergraphs for the 34-17 (left) and 26-13 (right) proofs shown in Table \ref{34_17}.}
\label{fig:3417-2613}
\end{figure}

A second example of a parity proof is given in Table
\ref{34_17}. This table shows a 40-30 set containing
a 34-17 and a 26-13 parity proof within it. MMP
hypergraph representations of these proofs are shown
in Figure \ref{fig:3417-2613}. The 34-17 proof has
a decagonal loop of bases for its  skeleton and
the 26-13 proof an octagonal loop, with the remaining
bases in both cases straddling different parts of
the skeleton. Constructions for the parity proofs
in Tables \ref{30_15} and \ref{34_17} based on the
geometry of the 600-cell are given in the Appendix.
Several other examples of parity proofs, together
with their constructions, can also be found there.

\begin{table}[htp]
\vskip-10pt
\addtolength{\tabcolsep}{-3pt}
\renewcommand{\arraystretch}{1.2}
\begin{center}
\begin{tabular}{|c|cccc|cccc|cccc|cccc|cccc|}
\hline
&\multicolumn{4}{|c|}{A}&\multicolumn{4}{|c|}{B}&
\multicolumn{4}{|c|}{C}&\multicolumn{4}{|c|}{D}&\multicolumn{4}{|c|}{E}\\
\hline
\multirow{3}{*}{A'}&&&&&&&&&\bf 22&\bf 60&\bf 39&\bf 28&\bf 57&\bf
23&\bf 27&\bf 40&&&&\\[-3pt]
&\bf 5&\bf 6&\bf 7&\bf 8&38&24&58&25&&&&&&&&&59&26&37&21\\[-3pt]
&\bf\underline 9&\bf\underline{10}&\bf\underline{11}&
\bf\underline{12}&&&&&&&&&&&&&&&&\\
\hline
\multirow{3}{*}{B'}&&&&&&&&&\textbf{\em 34}&\textbf{\em 12}&
\textbf{\em 51}&\textbf{\em 40}&\textbf{\em 9}&\textbf{\em 35}&
\textbf{\em 39}&\textbf{\em 52}&&&&\\[-3pt]
&17&18&19&20&\bf\underline{50}&\bf\underline{36}&
\bf\underline{10}&\bf\underline{37}&&&&&&&&&\bf
\underline{11}&\bf\underline{38}&\bf\underline{49}&\bf\underline{33}\\[-3pt]
&\bf 21&\bf 22&\bf 23&\bf 24&&&&&&&&&&&&&&&&\\
\hline
\multirow{3}{*}{C'}&25&26&27&28&&&&&&&&&&&&&&&&\\[-3pt]
&&&&&&&&&42&11&57&19&60&17&44&10&&&&\\[-3pt]
&\bf\underline{33}&\bf\underline{34}&\bf\underline{35}&
\bf\underline{36}&\bf 20&\bf 9&\bf 41&\bf 59&&&&&&&&&\bf
58&\bf 43&\bf 12&\bf 18\\
\hline
\multirow{3}{*}{D'}&\bf\underline{37}&\bf\underline{38}&
\bf\underline{39}&\bf\underline{40}&\bf 7&\bf 18&\bf 27&\bf
52&&&&&&&&&\bf 20&\bf 5&\bf 51&\bf 28\\[-3pt]
&41&42&43&44&&&&&&&&&&&&&&&&\\[-3pt]
&&&&&&&&&49&8&26&17&19&25&6&50&&&&\\
\hline
\multirow{3}{*}{E'}&\bf\underline{49}&\bf\underline{50}&
\bf\underline{51}&\bf\underline{52}&&&&&&&&&&&&&&&&\\[-3pt]
&&&&&&&&&\bf 6&\bf 35&\bf 21&\bf 43&\bf 24&\bf 41&\bf 8&\bf 34&&&&\\[-3pt]
&\bf 57&\bf 58&\bf 59&\bf 60&44&33&5&23&&&&&&&&&22&7&36&42\\
\hline
\end{tabular}
\end{center}
\caption{All 30 bases shown are those of a 40-30
set, and all the bold bases (plain bold, bold italic
and bold underlined) are those of a 34-19 set. The
two parity proofs are provided by a 34-17 set (whose
bases are the plain bold and bold underlined ones)
and a 26-13 set (whose bases are the plain bold
and bold italic ones)}.
\label{34_17}
\vskip-20pt
\end{table}

\eject

\section{\label{sec:discussion}Discussion}

      We have pointed out the existence of a large
number of $R$-$B$ (i.e., Ray-Basis) sets within
the 600-cell that provide parity proofs of the BKS
theorem. Some general observations can be made about these sets:
\begin{enumerate}[(i)]
\item All the sets listed in Table \ref{Overview}
are basis-critical (as defined in Sec.~\ref{sec:parity-proof}).
We checked this by means of a computer program.
\item Although Table \ref{Overview} lists only 90
critical sets, the actual number is much larger,
for two reasons. The first reason is that many of
the sets in Table \ref{Overview} come in a number
of distinct varieties that are not equivalent to
each other under the symmetries of the 600-cell.
One example of this is provided by the 30-15 sets,
of which there are six different varieties. In addition
to the two different varieties shown in Table \ref{30_15}
(which are really different, despite their structurally
identical MMP diagrams), a third type is shown in
Table \ref{38_19} and there are three further types
that we have not exhibited here. The differences
between these types can be brought out by calculating
the inner products of vectors in each of them, whereupon
it will be found that the patterns of the inner
products are not the same. These differences are
experimentally significant, because the unitary
transformations needed to transform the standard
basis into the bases of each of these types are
different. A few sets, such as the 26-13 set, come
in only one variety, but the vast majority come
in a number of different varieties. The second reason
is that each of the geometrically distinct critical
sets for a particular set of $R$ and $B$ values
has many replicas (typically in the thousands) under
the symmetries of the 600-cell. The combined effect
of both these factors is to increase the total number
of distinct parity proofs to somewhere in the vicinity
of a hundred million.
\item We have limited our discussion in this paper
only to critical sets that provide parity proofs
of the BKS theorem. However the 600-cell has a large
number of critical sets that provide non-parity
proofs of the theorem. These proofs are not as transparent
as the parity proofs, but they are just as conclusive.
We explored them in part in Ref.~\cite{Pavicic2010}
and will analyze and generate them extensively in
Ref.~\cite{Aravind2010}.
\item The parity proofs of this paper can be used
to devise experimental tests of noncontextuality
of the sort proposed by Cabello \cite{Cabello2008}.
We recall how such a test works. For a $R$-$B$ set
yielding a parity proof, let $A_j^i = 2\left| {\psi
_j^i} \right\rangle \left\langle {\psi _j^i} \right|
- 1$ $\left( {i = 1, \cdots ,B,j = 1, \cdots ,4}
\right)$, where $\left| {\psi _j^i} \right\rangle$
is the normalized column vector corresponding to
the $j-$th ray of the $i-$th basis (note that two
or more of the $\psi_j^i$  with different values
of $i$ and/or $j$ can be identical because the same
ray generally occurs in several different bases).
Each observable $A_j^i$ has only the eigenvalues
$+1$ or $-1$.  Cabello's argument implies that any
noncontextual hidden variables theory (NHVT) obeys the
inequality

\begin{equation}
\sum\limits_{i = 1}^B { - \left\langle
{A_1^iA_2^iA_3^iA_4^i} \right\rangle }  \le M  ,
\label{parityarg}
\end{equation}
where the averages $\left\langle \right\rangle$
above are to be taken over an ensemble of runs and
$M$ is an upper bound. Quantum mechanics predicts
that (\ref{parityarg}) holds as an equality with
$M = B$, but NHVTs predict
(see next paragraph) that the above inequality holds
with $M$ equal to $B-2$ at most. This is the contradiction
between NHVTs and quantum mechanics that
can be put to experimental test.

\medskip
We now give the argument leading to the maximum
value of $M$, namely, $B-2$. According to a NHVT,
each observable $A_j^i$ has the definite value of
+1 or -1 in any quantum state, independent of the
other observables with which it is measured.
Consider the expression on the left side of (\ref{parityarg}),
but without the averaging $\left\langle \right\rangle$
over many runs, and denote it by $F$. The maximum
value of $F$ in any run is $B$, and it is achieved
when each term in it has the value of +1. Let us
see how the values of the various $A_j^i$ can be
chosen so that this maximum is achieved. Clearly,
one or three of the $A_j^i$'s in each term of $F$
must be equal to -1 for this to happen. Let the
number of terms with one $A_j^i$ equal to -1 be
$n$ and the number with three $A_j^i$'s equal to
-1 be $m$. Then, if we choose the values of the
$A_j^i$'s in such a way that $n + m = B$, we can
guarantee that $F=B$. But an obstacle looms that
prevents us from reaching this goal. The total number
of -1's occurring over all the bases is $n + 3m
= B + 2m$ (since $n + m = B$). The difficulty now
is that $B + 2m$ is required to be both odd and
even (odd because $B$ is odd, and even because the
number of -1's in all the bases is required to be
even for the parity proof to be valid). This contradiction
shows that a NHVT cannot make the
value of $F$ equal to $B$. The best it can do is
to make all but one of the terms in $F$ equal to
+1, and this limits the maximum value of $F$ to
$B-2$. Averaging the value of $F$ over a large number
of runs could make the quantity on the left side
of (\ref{parityarg}) dip below the upper bound of
$B-2$, according to a NHVT. 

\medskip
For any basis-critical parity proof, quantum mechanics predicts that (\ref{parityarg}) holds as an equality with $M = B$ whereas a NHVT predicts that $M = B-2$ (since value assignments can always be found that make $B - 1$ of the terms on the left of (\ref{parityarg}) equal to 1 and one term equal to -1). A 18-9 parity proof thus leads to the ratio of 7/9 for the bounds due to NHVTs and quantum mechanics. This bound can be improved slightly by considering all 1800 26-13 parity proofs within the 60-75 set. Of the 75 terms on the left side of (\ref{parityarg}), at least 9 must then contribute -1 to the sum, causing the previous ratio to dip to 57/75, which is very slightly less than 7/9. Whether a further improvement can be effected by simultaneous consideration of a larger number of parity proofs is an open question.

\medskip
It is worth stressing that the contradiction we
have demonstrated between NHVTs and quantum
mechanics generalizes in a straightforward manner to any parity proof
in any even dimension greater than or equal to 4. Parity proofs of the type
we are considering are not possible in odd dimensions, so a similar conflict
cannot be demonstrated in this case. 
\item Any parity proof of the BKS theorem (or even
a non-parity proof) can be turned into a scheme
for quantum key distribution, as pointed out in
\cite{BPPeres}. The idea is simple: since there
are no hidden variables that model the observables
in a BKS proof, there is no data in the transmitted
particles to be stolen while the key is being established;
the key comes into being only after sender and receiver
exchange messages to determine the cases in which
they used the same bases to encode and decode their
particles. The preferred bases in such a scheme,
when they exist, are a maximal set of mutually unbiased
bases. A maximal set of five mutually unbiased bases
does indeed exist in four dimensions, and has been
proposed for use in key distribution schemes based
on four-state systems \cite{GisinRMP}. However,
any set of bases leading to a BKS proof, such as
the ones in this paper, can also be used. They may
not be as efficient as schemes based on mutually
unbiased bases, but they may be advantageous in
some situations and would therefore seem to be worth
exploring further.
\end{enumerate}

\appendix
\section{\label{sec:appendix}Appendix}
The purpose of this Appendix is to show how special
geometrical features of the 600-cell can be exploited
to give simple rules for generating many of the
parity proofs in Table \ref{Overview}. We first
review some basic geometrical facts about the 600-cell
and then show how they can be used to arrive at
the rules. Readers wanting a more detailed account
of the geometrical properties of the 600-cell can
consult the classic monograph by Coxeter \cite{Coxeter}.

The 600-cell is a regular polytope with 120 vertices
distributed symmetrically on the surface of a four-dimensional
sphere. The vertices come in antipodal pairs, and
the 60 rays are the unoriented directions passing
through antipodal pairs of vertices. If the vertices
are taken to lie on a sphere of radius 2 centered
at the origin, then the coordinates of 60 of the
vertices can be chosen as in Table \ref{tab:Ray}
and the remaining vertices are the antipodes of these.

\subsection{\label{subsec:600-facets}Facets of the 600-cell}
\begin{figure}[htp]
\begin{center}
\includegraphics[width=0.9\textwidth]{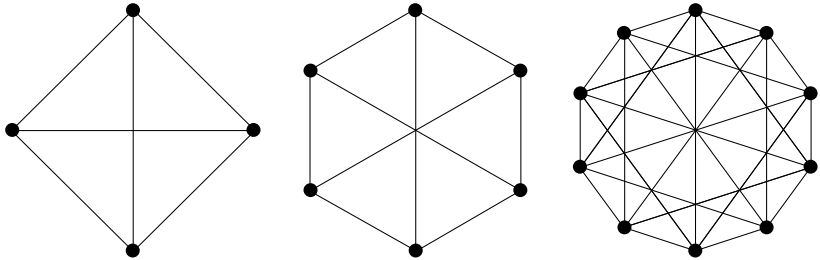}
\end{center}
\caption{The Kochen-Specker diagrams of a basis (left), a dual line
pair (DLP) (center) and a dual pentagon pair (DPP) (right).}
\label{KSdiag}
\end{figure}

\vskip-5pt
Any finite subset of rays of the 600-cell will be
termed a {\bf facet}. There are several facets that
will play an important role in the constructions below.
We now define these facets and depict them by their
Kochen-Specker (KS) diagrams in Figure \ref{KSdiag}.
In a KS diagram, rays are represented by dots and dots
corresponding to orthogonal pairs of rays are joined
by lines.

The simplest type of facet is a single ray, which we
will also term a {\bf point}. Rays/points will be
referred to by the numbers assigned to them in Table
\ref{tab:Ray}.

Four mutually orthogonal rays make up a {\bf basis}.
The KS diagram of a basis can be taken as four dots
at the corners of a square, with all the edges and
diagonals of the square drawn in. A basis will be
denoted by four numbers separated by spaces, e.g.,
7 18 27 52, just as in Table \ref{tab:Basis}.

Any three rays will be said to form a {\bf line}
if the coordinates of one of them can be expressed
as a linear combination of those of the other two.
 For example, the points 3,8 and 10 form a line,
which we will denote (3 8 10). Two lines are said
to be the duals of each other if every point (ray)
on one is orthogonal to every point on the other.
The lines (3 8 10) and (16 17 24) are the duals
of each other and will be termed a {\bf dual line
pair} (DLP) and denoted (3 8 10)+(16 17 24). The
KS diagram of a DLP takes on a simple form if the
points of the dual lines are arranged at the alternate
vertices of a hexagon; then the orthogonalities
between the rays are represented by the six edges
and three diameters of the hexagon. The geometry
of the 600-cell is such that any set of six points
having this KS diagram represents a dual line pair;
the requirement that the points at the alternate
vertices satisfy the conditions for a line need
not be added because it turns out to be satisfied
automatically.
\vskip-15pt
\vskip-15pt

A {\bf pentagon} is any set of five rays with the
property that five pairs among them have an absolute
inner product of $\tau$ and the remaining five an
absolute inner product of $\kappa$. The points 1,15,30,47
and 56 form a pentagon, which we will denote (1
15 30 47 56).  Two pentagons will be said to be
the duals of one another if every point of one is
orthogonal to every point of the other. The pentagons
(1 15 30 47 56) and (4 14 29 45 55) are the duals
of each other and will be said to be a {\bf dual
pentagon pair} (DPP). A symbol for a DPP will
be introduced in the next subsection. If the points
corresponding to a dual pair of pentagons are arranged at the alternate
vertices of a decagon, then the KS diagram of the
DPP takes on a very simple form: it consists of
the ten edges and five diameters of the decagon,
together with ten of its diagonals (see Figure \ref{KSdiag}).
Again, it turns out that any set of ten points possessing
this KS diagram constitutes a DPP, with the conditions
for the points at alternate vertices to form pentagons
being automatically satisfied.

The most complex facet of interest to us is a {\bf
Reye's configuration} (RC), which is a set of 12
``points'' and 16 ``lines'' with the property that
three ``points'' lie on every ``line'' and four
``lines'' pass through every ``point''. If the terms
``points'' and ``lines'' in this definition are
taken to be identical with the points and lines
defined above, it is easy to check that points 1
through 12 form a RC with the 16 lines being given
by (1 5 9), (1 6 10), (1 7 11),  (1 8 12), (2 5
10), (2 6 9), (2 7 12), (2 8 11), (3 5 11), (3 6
12), (3 7 9), (3 8 10), (4 5 12), (4 6 11), (4 7
10) and (4 8 9). An equivalent definition of a RC
is that it consists of the rays in three mutually
unbiased bases (i.e. bases with the property that
the magnitude of the inner product of any normalized
ray of one with any normalized ray of the other
is always the same). In the case of the 600-cell
this latter definition guarantees that the 12 rays
in the three bases form 16 lines, with one point
of each line coming from each of the three bases.
In the example of the RC just given, the three (mutually
unbiased) bases are 1 2 3 4, 5 6 7 8 and 9 10 11 12.

The above discussion has been carried out in projective or ray
space. However, many of the facets correspond to familiar
figures in four-dimensional Euclidean space if one
recalls that each point in projective space corresponds to a pair of
mutually inverse points in Euclidean space. Then a basis corresponds
to a {\bf 16-cell} (or {\bf cross polytope}), a pair of mutually
unbiased bases to a {\bf 8-cell} (or {\bf hypercube}), and three
mutually unbiased bases (or a RC) to a {\bf 24-cell}. These three
figures are all convex regular polytopes in four dimensions (just like
the 600-cell), and their bounding cells consist of 16 tetrahedra, 8 cubes
and 24 octahedra, respectively.

\subsection{\label{subsec:600-tilings}Tilings of the 600-cell by its facets}
\vskip-5pt
A particular type of facet (e.g. bases) will be
said to tile the 600-cell if the union of several
mutually disjoint specimens of that type yields
all 60 rays of the 600-cell. The 600-cell has many
tilings by its bases, DLPs, DPPs and RCs. We now
discuss these tilings one by one.

First consider the bases. The 600-cell has 75 bases
in it, which are shown in Table \ref{tab:Basis}.
The three bases in any block make up a RC, and the
5 x 5 array of blocks shows the 25 different RCs
in the 600-cell. The five RCs in any row or column
of the array give a tiling of the 600-cell. There
are exactly ten such tilings, one associated with
each row or column of the array. These tilings were
first discovered by the Dutch geometer P.H.Schoute
\cite{Coxeter}, who observed that the 600-cell has
five mutually disjoint 24-cells inscribed in it
in ten different ways. The letters in Table \ref{tab:Basis}
help in reconstructing these tilings from the 24-cells:
if each RC (or 24-cell) is labeled by a pair of
letters, a primed one and an unprimed one, then
arranging the RCs so that all the RCs in a row (or
column) share a primed (or unprimed) letter reproduces
the tilings. The tilings by bases are a trivial
consequence of the tilings by RCs: each of the latter
gives rise to a tiling by 15 bases.

Next consider the DLPs. The 600-cell has 100 DLPs
in it and they are arranged in a 10 x 10 array in
Table \ref{DLP}. The ten DLPs in any row or column
of this array give a tiling of the 600-cell, there
being 20 such tilings in all. Finally consider the
DPPs. There are 36 DPPs in the 600-cell and they
are arranged in a 6 x 6 array in Table \ref{DPP}.
The six DPPs in any row or column of this array
give a tiling of the 600-cell, there being 12 such
tilings in all. For later reference we will label
the DPPs in Table \ref{DPP} from 1 to 36, proceeding
from left to right and top to bottom. Thus, the
DPP in the third row and second column will be referred
to as DPP14 and the one in the last row and fifth
column as DPP35.

The basis table, DLP table and DPP table are closely
related in several ways. As one example of this, we
show how the basis table can be reconstructed from any
row or column of the DPP table. Any two DPPs from the
same row or column of Table \ref{DPP} can be ``mated'' to
produce five bases, with half the rays of each basis
coming from each of the DPPs. Since there are 15
pairings of the DPPs in a row or column, each of which
gives rise to five bases, the total number of bases
that can be produced in this way is 75, which are
all the bases of the 600-cell. The rows and columns
of the DPP table allow the basis table to be recovered
in 12 different ways.
\vskip-55pt
\subsection{\label{subsec:ray-deletion}Deletion of rays; Isogonal subsets of
the 60-ray system}
\vskip-5pt
By deletion of a set of rays from a $R$-$B$ set, we
will mean dropping all bases involving any of these
rays from this set to obtain a new set, $R'$-$B'$, with $R'<R$
and $B'<B$. Deletion of rays is a crucial step in
the construction of all the critical sets to be
presented below.

The 600-cell, as well as the 60-75 system of rays
and bases derived from it, is isogonal (or vertex/ray-transitive)
in the sense that there are symmetry operations
that take any vertex (or ray) into any other
vertex (or ray) while keeping the structure as a
whole invariant. It turns out that the 60-75 set
has a number of isogonal subsets within it, which
may be obtained by deleting any number of DPPs (from
one to five) from an arbitrary  row or column of
Table \ref{DPP}. Deletion of one, two or three DPPs
from any row or column of Table \ref{DPP} from the 60-75 set reduces
it to a 50-50, 40-30 or 30-15 set, respectively.
The basis tables of these sets are decimated versions
of Table \ref{tab:Basis}, with 25,35 or 45 bases dropped,
respectively. The number of different 50-50, 40-30
and 30-15 sets is 36, 180 and 240, respectively.
These smaller isogonal sets are interesting because
they each contain a large number of critical sets
and can be searched far more easily for these sets
than the full 60-75 set.  We will ignore the
sets obtained by deleting four or more DPPs from
a row or column of Table \ref{DPP} because they
do not contain any critical sets.

\subsection{\label{subsec:critical-sets}Constructions for some critical sets}
\vskip-5pt
We now use the ideas and tools developed in the
previous subsections to give constructions for some
of the parity proofs in Table \ref{Overview}.

\subsubsection{\label{subsubsec:30-15}30-15 set(Type-1)}
\vskip-5pt
A somewhat involved construction for this set was
given in \cite{Waegell2010}, based on the deletion
of DLPs. However a much simpler procedure is to delete any three DPPs from 
the same row or column of Table \ref{DPP} from the 60-75 set. It was pointed out in Sec. \ref{subsec:ray-deletion}
that this procedure gives rise to 240 isogonal 30-15
sets. Inspection of these sets shows that they all
provide parity proofs of the BKS theorem! An interesting
feature of these sets is that they come in 120 complementary
pairs, with the members of each pair having no rays
in common. The two members of a pair are obtained
by deleting distinct triads of DPPs from the same
row or column of Table \ref{DPP}. For example, for
the complementary pairs shown in Table \ref{30_15},
the one in bold is obtained by deleting DPPs 1,2
and 3 and the one in plain type by deleting DPPs
4,5 and 6. Alternatively, the former set is obtained
by keeping all bases involving only the rays in
DPPs 4,5 and 6 and the latter by keeping all bases
involving only the rays in DPPs 1,2 and 3. These
sets have been termed Type-1 because they are geometrically
distinct from the Type-2 30-15 sets to be presented
a little later.

The MMP hypergraphs of Fig. \ref{fig:30-15-a-b}
have a nice interpretation in terms of DPPs. The
points at the vertices of each decagon represent
a DPP, with alternate vertices representing its
two component pentagons. The two points within each
edge in an alternating set of edges of the decagon
also represent a DPP, there being two such DPPs.
For each of the latter DPPs, the two points on an
edge come from dual pentagons, and again the points
alternate between the pentagons as one goes around
the loop. It was pointed out in Sec. \ref{subsec:600-tilings}
that any two DPPs mate to produce five bases, with
two rays in each of the bases coming from each of
the DPPs. This explains how the 15 bases arise in
each of the hypergraphs: the bases corresponding
to the edges arise from the matings between the
distinguished DPP (corresponding to the decagon
vertices) and each of the others, while the bases
that straddle the figure arise from the matings
of the other two DPPs with each other.

\subsubsection{\label{subsubsec:34-17}34-17 set}
\vskip-5pt
This set can be constructed as follows:
\begin{enumerate}[(i)]
\item Pick a 40-30 set by deleting any two DPPs from
the same row or column of Table \ref{DPP}. For example,
deleting DPPs 1 and 2 gives the 40-30 set shown
in Table \ref{34_17}.
\item Delete any DLP from this 40-30 set to get a
34-19 set.  For example, deleting the DLP (5 24
57)+(8 23 58) leads to the 34-19 set in Table \ref{34_17}
whose bases are shown in plain boldface, italic
boldface and underlined boldface.
\item The 34-19 set has 8 rays that each occur thrice
in it and 26 rays that each occur twice.  The 8
rays that each occur thrice form two bases made
up of just themselves; these are the italic boldface
bases in Table \ref{34_17}. Dropping these bases
from the 34-19 set gives a 34-17 set.
\end{enumerate}
     The number of 34-17 sets that can be constructed
in this way is the product of the number of 40-30
sets that can be picked in step (i)(= 180) and the
number of DLPs that can be deleted in step (ii)(=
20), or 3600.

\subsubsection{\label{subsubsec:26-13}26-13 set}
\vskip-5pt
A 26-13 set can be constructed by modifying
the above procedure slightly. One keeps steps (i)
and (ii), but replaces (iii) by the following alternative
step:
\begin{enumerate}[(iii')]
\item Write the two italic boldface bases in Table \ref{34_17}
(that were dropped in getting the 34-17 set) horizontally, one below
the other, in such a way that each vertical pair of rays can be
augmented by two additional rays to form a basis. This is done
below, with the eight added rays indicated in boldface.
The eight added rays (which are always unique) lead
to six new bases, four along the columns of the
array and two more along its last two rows. These
six new bases are all present in Table \ref{34_17}
and are the underlined boldfaced ones. Dropping
these bases from the 34-19 set gives a 26-13 set.
\end{enumerate}

\begin{table}[htp]
\centering 
\begin{tabular}{c c c c} 
9 &  35  & 39  & 52 \\
 12  & 34  & 40  & 51 \\
{\bf  10} & {\bf 36}  &{\bf 37}  & {\bf 50} \\
 {\bf11}  & {\bf33}  & {\bf38} &  {\bf 49} \\
\end{tabular}
\vskip-20pt
\end{table}
It might appear that the number of 26-13 sets that
can be constructed in this way is the same as the
number of 34-17 sets, or 3600. However it turns
out that every 26-13 set is obtained twice by this
method, so that their true number is 1800. As an
illustration of this, the 26-13 set in Table \ref{34_17}
can also be constructed by first deleting DPPs 21
and 24, then deleting the DLP (3 19 56)+(4 17 54)
and finally truncating the resulting 34-19 set in
the manner described in step (iii').

\subsubsection{\label{subsubsec:38-19}38-19 set}
\vskip-5pt
The procedure for constructing this set is as follows:
\begin{enumerate}
\item Choose a 50-50 set by deleting an arbitrary
DPP. For example, deleting DPP1 leads to the 50-50
set shown in Table \ref{38_19}.
\begin{table}[htp]
\vskip-10pt
\addtolength{\tabcolsep}{-3pt}
\begin{center}
\begin{tabular}{|c|cccc|cccc|cccc|cccc|cccc|}
\hline
&\multicolumn{4}{|c|}{A}&\multicolumn{4}{|c|}{B}&
\multicolumn{4}{|c|}{C}&\multicolumn{4}{|c|}{D}&
\multicolumn{4}{|c|}{E}\\
\hline
\multirow{3}{*}{A\'}&&&&&\textbf{31}&
\textbf{42}&\textbf{51}&\textbf{16}&
\underline{\textbf{22}}&\underline{\textbf{60}}&
\underline{\textbf{39}}&\underline{\textbf{28}}&
\underline{\textbf{57}}&\underline{\textbf{23}}&
\underline{\textbf{27}}&\underline{\textbf{40}}&&&&\\
&5&6&7&8&\underline{\textbf{38}}&
\underline{\textbf{24}}&\underline{\textbf{58}}&
\underline{\textbf{25}}&&&&&36&53&20&46&
\underline{\textbf{59}}&\underline{\textbf{26}}&
\underline{\textbf{37}}&\underline{\textbf{21}}\\
&\textbf{9}&\textbf{10}&\textbf{11}&\textbf{12}&&&&&
\textbf{13}&\textbf{32}&\textbf{50}&\textbf{41}&&&&&
34&19&48&54\\
\hline
\multirow{3}{*}{B\'}&&&&&43&54&3&28&34&12&51&40&
\textbf{9}&\textbf{35}&\textbf{39}&\textbf{52}&&&&\\
&17&18&19&20&50&36&10&37&&&&&48&5&32&58&\textbf{11}&
\textbf{38}&\textbf{49}&\textbf{33}\\
&\underline{\textbf{21}}&\underline{\textbf{22}}&
\underline{\textbf{23}}&\underline{\textbf{24}}&&&&&
25&44&2&53&&&&&46&31&60&6\\
\hline
\multirow{3}{*}{C\'}&\underline{\textbf{25}}&
\underline{\textbf{26}}&\underline{\textbf{27}}&
\underline{\textbf{28}}&&&&&46&24&3&52&&&&&8&53&39&16\\
&&&&&2&48&22&49&42&11&57&19&\textbf{60}&\textbf{17}&
\textbf{44}&\textbf{10}&&&&\\
&33&34&35&36&20&9&41&59&&&&&7&13&54&38&\textbf{58}&
\textbf{43}&\textbf{12}&\textbf{18}\\
\hline
\multirow{3}{*}{D\'}&\textit{\textbf{37}}&
\textit{\textbf{38}}&\textit{\textbf{39}}&
\textit{\textbf{40}}&7&18&27&52&&&&&\textbf{33}&
\textbf{59}&\textbf{3}&\textbf{16}&20&5&51&28\\
&\textbf{41}&\textbf{42}&\textbf{43}&\textbf{44}&&&&&
54&23&9&31&&&&&\textbf{35}&\textbf{2}&\textbf{13}&\textbf{57}\\
&&&&&32&21&53&11&49&8&26&17&19&25&6&50&&&&\\
\hline
\multirow{3}{*}{E\'}&\textbf{49}&\textbf{50}&\textbf{51}&
\textbf{52}&&&&&10&48&27&16&&&&&\textbf{32}&
\textbf{17}&\textbf{3}&\textbf{40}\\
&&&&&26&12&46&13&6&35&21&43&24&41&8&34&&&&\\
&\textit{\textbf{57}}&\textit{\textbf{58}}&
\textit{\textbf{59}}&\textit{\textbf{60}}&44&33&5&23&&&&&
\textbf{31}&\textbf{37}&\textbf{18}&\textbf{2}&22&7&36&42\\
\hline
\end{tabular}
\end{center}
\caption{All 50 bases are those of a 50-50 set and
all the bold bases (plain bold, bold italic and
bold underlined) are those of a 38-21 set. The two
parity proofs are provided by a 38-19 set (whose
bases are the plain bold and bold underlined ones)
and a 30-15 set (whose bases are the plain bold
and bold italic ones)}.
\label{38_19}
\vskip-20pt
\end{table}
\item This 50-50
set (like all 50-50 sets) has 50 DLPs in it. Define
the separation of two DLPs as the number of orthogonalities
of rays between the two. It turns out that any 50-50
set has exactly 100 pairs of DLPs of separation
12.  Deleting any such pair from a 50-50 set will
lead to a 38-21 set. In the example of Table \ref{38_19},
deleting the DLPs (5 19 46)+(6 20 48) and (7 34
53)+(8 36 54) leads to the 38-21 set whose bases
are the ones shown in plain boldface, italic boldface
and underlined boldface.
\item The 38-21 set has 8 rays that each occur thrice
in it and 30 rays  that each occur twice.  The 8 rays
that each occur thrice form two bases made up of just
themselves; these are the italic boldface bases in
Table \ref{38_19}. Dropping these bases from the 38-21
set gives a 38-19 set.
\end{enumerate}

The number of 38-19 sets that can be constructed by this
method is the product of the number of 50-50 sets that
can be picked in the first step (= 36) and the number of
pairs of DLPs of separation 12 that can be deleted in
the second step (= 100), or 3600.

\subsubsection{\label{subsubsec:30-15-type2}30-15 set (Type 2)}
\vskip-5pt
This set can be obtained from a 38-21 set in a manner
similar to that in which a 26-13 set is obtained
from a 34-17 set. After carrying out steps (i) and
(ii) for the 38-19 set just discussed, replace step
(iii) by the following alternative step: \\
(iii') Write the two italic boldface bases in Table
\ref{38_19} (that were dropped in getting the 38-19
set) horizontally, one below the other, in such
a way that each vertical pair of rays can be augmented
by two additional rays to form a basis. This is
shown below, with the eight added rays indicated in boldface.

\begin{table}[htp]
\vskip-15pt
\centering 
\begin{tabular}{c c c c} 
37 &  38  & 39  & 40 \\
 59  & 58  & 60  & 57 \\
{\bf  21} & {\bf 24}  &{\bf 22}  & {\bf23} \\
 {\bf26}  & {\bf25}  & {\bf28} &  {\bf 27} \\
\end{tabular}
\vskip-15pt
\end{table}

The eight added rays (which are unique) lead to
six new bases, four along the columns of the array
and two more along its last two rows. These six
new bases are all present in Table \ref{38_19} and
are the underlined boldfaced ones. Dropping these
bases from the 38-21 set gives a Type-2 30-15 set.
The number of such sets is the same as the number
of 38-21 sets, or 3600. Note that we have called
this 30-15 set a Type-2 set to distinguish it from
the one constructed in Sec.\ref{subsubsec:30-15}
These two types of 30-15 sets are geometrically
distinct in that the rays of one cannot be made
to pass into those of the other by any rotation
in four-dimensional space.

\subsubsection{\label{subsubsec:50-25}50-25 set}
\vskip-5pt
This set can be constructed by deleting an arbitrary
DPP from the 60-75 set to obtain a 50-50 set and
then dividing the latter (in 291 ways) into a pair
of 50-25 sets. The two 50-25 sets that are obtained
in this way involve the same 50 rays but have no
bases in common. The sets shown in Table \ref{50_25}
were obtained by deleting DPP10 and then partitioning
the bases.
\begin{table}[htp]
\vskip-20pt
\addtolength{\tabcolsep}{-2.9pt}
\begin{center}
\begin{tabular}{|c|cccc|cccc|cccc|cccc|cccc|}
\hline
&\multicolumn{4}{|c|}{A}&\multicolumn{4}{|c|}{B}&
\multicolumn{4}{|c|}{C}&\multicolumn{4}{|c|}{D}&
\multicolumn{4}{|c|}{E}\\
\hline
\multirow{3}{*}{A\'}&\textbf{1}&\textbf{2}&\textbf{3}&
\textbf{4}&\textbf{31}&\textbf{42}&\textbf{51}&
\textbf{16}& & & & & & & & &\textbf{44}&\textbf{29}&
\textbf{15}&\textbf{52}\\
& & & & &\textbf{38}&\textbf{24}&\textbf{58}&
\textbf{25}&\textbf{18}&\textbf{47}&\textbf{33}&
\textbf{55}&\textbf{36}&\textbf{53}&\textbf{20}&
\textbf{46}&\textbf{59}&\textbf{26}&\textbf{37}&\textbf{21}\\
&\textbf{9}&\textbf{10}&\textbf{11}&\textbf{12}& & & & &
\textbf{13}&\textbf{32}&\textbf{50}&\textbf{41}&
\textbf{43}&\textbf{49}&\textbf{30}&\textbf{14}& & & & \\
\hline
\multirow{3}{*}{B\'}&13&14&15&16& & & & &\textbf{34}&
\textbf{12}&\textbf{51}&\textbf{40}&\textbf{9}&
\textbf{35}&\textbf{39}&\textbf{52}& & & & \\
&\textbf{17}&\textbf{18}&\textbf{19}&\textbf{20}&
50&36&10&37& & & & & & & & &11&38&49&33\\
& & & & &\textbf{8}&\textbf{57}&\textbf{29}&
\textbf{47}&25&44&2&53&55&1&42&26&\textbf{46}&
\textbf{31}&\textbf{60}&\textbf{6}\\
\hline
\multirow{3}{*}{C\'}& & & & &55&6&15&40&46&24&3&
52&21&47&51&4&8&53&39&16\\
&29&30&31&32& & & & &42&11&57&19&60&17&44&10& & & & \\
&33&34&35&36&20&9&41&59& & & & & & & & &58&43&12&18\\
\hline
\multirow{3}{*}{D\'}&\textbf{37}&\textbf{38}&
\textbf{39}&\textbf{40}& & & & &\textbf{58}&
\textbf{36}&\textbf{15}&\textbf{4}&\textbf{33}&
\textbf{59}&\textbf{3}&\textbf{16}& & & & \\
&\textbf{41}&\textbf{42}&\textbf{43}&\textbf{44}&
\textbf{14}&\textbf{60}&\textbf{34}&\textbf{1}& & &
& & & & & &\textbf{35}&\textbf{2}&\textbf{13}&\textbf{57}\\
& & & & &\textbf{32}&\textbf{21}&\textbf{53}&
\textbf{11}&\textbf{49}&\textbf{8}&\textbf{26}&
\textbf{17}&\textbf{19}&\textbf{25}&\textbf{6}&
\textbf{50}&\textbf{10}&\textbf{55}&\textbf{24}&\textbf{30}\\
\hline
\multirow{3}{*}{E\'}&49&50&51&52&19&30&39&4& & &
& & & & & &32&17&3&40\\
& & & & &26&12&46&13&6&35&21&43&24&41&8&34&47&14&25&9\\
&57&58&59&60& & & & &1&20&38&29&31&37&18&2& & & & \\
\hline
\end{tabular}
\end{center}
\vskip-5pt
\caption{Two 50-25 parity proofs, one in plain type
and the other in boldface, both involving the same
50 rays but having no bases in common.}
\vskip-20pt
\label{50_25}
\end{table}

The total number of 50-25 sets that can
be constructed in this way is the product of the
number of 50-50 sets that can be picked in the first
step (= 36) and the number of 50-25 sets into which
each can be divided (= 2x291), or 20,952.

\subsubsection{\label{subsubsec:54-27}54-27 set}
\vskip-5pt
The construction of this set is similar to the last,
but with a small twist. This time one deletes an
arbitrary DLP from the 60-75 set to get a 54-54
set and then divides the latter (in 368 ways) into
two 54-27 sets. The sets shown in Table \ref{54_27}
were obtained by deleting the DLP (5,19,46)+(6,20,48)
and then partitioning the bases. The number of 54-27
sets that can be constructed in this way is the
product of the number of 54-54 sets that can be
picked in the first step (= 100) and the number
of 54-27 sets into which each can be divided (=
2x368), or 73,600.

\begin{table}[htp]
\vskip-10pt
\addtolength{\tabcolsep}{-2.9pt}
\begin{center}
\begin{tabular}{|c|cccc|cccc|cccc|cccc|cccc|}
\hline
&\multicolumn{4}{|c|}{A}&\multicolumn{4}{|c|}{B}&
\multicolumn{4}{|c|}{C}&\multicolumn{4}{|c|}{D}&
\multicolumn{4}{|c|}{E}\\
\hline
\multirow{3}{*}{A\'}&\textbf{1}&\textbf{2}&
\textbf{3}&\textbf{4}&\textbf{31}&\textbf{42}&
\textbf{51}&\textbf{16}&\textbf{22}&\textbf{60}&
\textbf{39}&\textbf{28}&57&23&27&40&\textbf{44}&
\textbf{29}&\textbf{15}&\textbf{52}\\[-3pt]
&&&&&\textbf{38}&\textbf{24}&\textbf{58}&
\textbf{25}&\textbf{18}&\textbf{47}&\textbf{33}&
\textbf{55}&&&&&59&26&37&21\\[-3pt]
&\textbf{9}&\textbf{10}&\textbf{11}&\textbf{12}&
\textbf{56}&\textbf{45}&\textbf{17}&\textbf{35}&
\textbf{13}&\textbf{32}&\textbf{50}&\textbf{41}&
\textbf{43}&\textbf{49}&\textbf{30}&\textbf{14}&&&&\\
\hline
\multirow{3}{*}{B\'}&13&14&15&16&43&54&3&28&
\textbf{34}&\textbf{12}&\textbf{51}&\textbf{40}&9&35&39&
52&56&41&27&4\\[-3pt]
&&&&&\textbf{50}&\textbf{36}&\textbf{10}&
\textbf{37}&\textbf{30}&\textbf{59}&\textbf{45}&
\textbf{7}&&&&&11&38&49&33\\[-3pt]
&\textbf{21}&\textbf{22}&\textbf{23}&\textbf{24}&
\textbf{8}&\textbf{57}&\textbf{29}&\textbf{47}&25&44&
2&53&55&1&42&26&&&&\\
\hline
\multirow{3}{*}{C\'}&\textbf{25}&\textbf{26}&
\textbf{27}&\textbf{28}&&&&&&&&&21&47&51&4&8&53&39&16\\[-3pt]
&29&30&31&32&&&&&&&&&60&17&44&10&23&50&1&45\\[-3pt]
&33&34&35&36&&&&&&&&&7&13&54&38&58&43&12&18\\
\hline
\multirow{3}{*}{D\'}&\textbf{37}&\textbf{38}&\textbf{39}&
\textbf{40}&\textbf{7}&\textbf{18}&\textbf{27}&\textbf{52}&
\textbf{58}&\textbf{36}&\textbf{15}&\textbf{4}&\textbf{33}&
\textbf{59}&\textbf{3}&\textbf{16}&&&&\\[-3pt]
&\textbf{41}&\textbf{42}&\textbf{43}&\textbf{44}&\textbf{14}&
\textbf{60}&\textbf{34}&\textbf{1}&\textbf{54}&\textbf{23}&
\textbf{9}&\textbf{31}&12&29&56&22&\textbf{35}&\textbf{2}&
\textbf{13}&\textbf{57}\\[-3pt]
&&&&&\textbf{32}&\textbf{21}&\textbf{53}&\textbf{11}&
\textbf{49}&\textbf{8}&\textbf{26}&\textbf{17}&&&&&10&55&24&30\\
\hline
\multirow{3}{*}{E\'}&49&50&51&52&&&&&&&&&45&11&15&28&32&17&3&40\\[-3pt]
&\textbf{53}&\textbf{54}&\textbf{55}&\textbf{56}&&&&&&&&&24&
41&8&34&47&14&25&9\\[-3pt]
&57&58&59&60&&&&&&&&&31&37&18&2&22&7&36&42\\
\hline
\end{tabular}
\end{center}
\vskip-7pt
\caption{Two 54-27 parity proofs, one in plain type and
the other in boldface, both involving the same 54 rays
but having no bases in common.}
\vskip-15pt
\label{54_27}
\end{table}

\subsubsection{\label{subsubsec:36-19}36-19 and 32-17 sets}
\vskip-5pt
In all the above proofs, each ray occurred twice among
the bases. Table \ref{36_19} shows two parity proofs,
a 36-19 proof and a 32-17 proof, that both involve two
rays occurring four times each.

\begin{table}[htp]
\vskip-15pt
\addtolength{\tabcolsep}{-2.93pt}
\begin{center}
\begin{tabular}{|c|cccc|cccc|cccc|cccc|cccc|}
\hline
&\multicolumn{4}{|c|}{A}&\multicolumn{4}{|c|}{B}&
\multicolumn{4}{|c|}{C}&\multicolumn{4}{|c|}{D}&
\multicolumn{4}{|c|}{E}\\
\hline
\multirow{3}{*}{A\'}&1&2&3&4&&&&&&&&&&&&&44&29&15&52\\[-3pt]
&&&&&38&24&58&25&&&&&\textbf{36}&\textbf{53}&
\textbf{20}&\textbf{46}&&&&\\[-3pt]
&&&&&&&&&13&32&50&41&&&&&&&&\\
\hline
\multirow{3}{*}{B\'}&13&14&15&16&&&&&&&&&
\textbf{9}&\textbf{35}&\textbf{39}&\textbf{52}&&&&\\[-3pt]
&&&&&&&&&&&&&&&&&&&&\\[-3pt]
&&&&&8&57&29&47&25&44&2&53&&&&&46&31&60&6\\
\hline
\multirow{3}{*}{C\'}&&&&&&&&&\textit{\textbf{46}}&
\textit{\textbf{24}}&\textit{\textbf{3}}&\textit{\textbf{52}}&
&&&&\textit{\textbf{8}}&\textit{\textbf{53}}&
\textit{\textbf{39}}&\textit{\textbf{16}}\\[-3pt]
&29&30&31&32&&&&&&&&&&&&&&&&\\[-3pt]
&\textbf{33}&\textbf{34}&\textbf{35}&\textbf{36}&
\textit{\textbf{20}}&\textit{\textbf{9}}&\textit{\textbf{41}}&
\textit{\textbf{59}}&&&&&&&&&&&&\\
\hline
\multirow{3}{*}{D\'}&&&&&&&&&&&&&\textbf{33}&
\textbf{59}&\textbf{3}&\textbf{16}&&&&\\[-3pt]
&&&&&&&&&&&&&&&&&&&&\\[-3pt]
&&&&&&&&&&&&&19&25&6&50&&&&\\
\hline
\multirow{3}{*}{E\'}&&&&&19&30&39&4&&&&&&&&&&&&\\[-3pt]
&&&&&&&&&&&&&\textbf{24}&\textbf{41}&\textbf{8}&
\textbf{34}&47&14&25&9\\[-3pt]
&57&58&59&60&&&&&1&20&38&29&&&&&&&&\\
\hline
\end{tabular}
\end{center}
\vskip-5pt
\caption{A 36-19 parity proof (bases in plain type
and bold) and a 32-17 parity proof (bases in plain
type and bold italics). Each proof has two rays
that occur four times each (rays 25 and 29 in both
cases), with all the remaining rays occurring twice
each.}
\vskip-15pt
\label{36_19}
\end{table}

\subsubsection{\label{subsubsec:BasisComp}Basis-complementary parity proofs}
\vskip-5pt
The proofs in Tables \ref{50_25} and \ref{54_27} are particular instances of a phenomenon we term "basis complementarity", which is of interest because it forges a link between many pairs of proofs in Table \ref{Overview} that might otherwise appear to be unrelated. Any $R$-$B$ set, with $B$ even, in which each ray occurs four times can potentially house many basis-complementary parity proofs within it. The 50-50 sets obtained by a deleting a DPP and the 54-54 sets obtained by deleting a DLP are both examples of such sets. If there is a $R$-$B$ parity proof contained in either of these sets, the remaining $N-B$ bases, which involve $N-2B+R$ distinct rays (with $N$ = 50 or 54), automatically yield another parity proof that we will term the "basis-complementary" proof to the original one.  The basis-complementary proof has $2R-2B$ rays that occur twice each and $N-R$ rays that occur four times each. In general the proof complementary to a given proof might not be basis-critical, in which case it would be left out of Table \ref{Overview}. Only if both members of a basis-complementary pair are basis-critical would they both be included in Table \ref{Overview} . Three examples of basis-complementary (and basis-critical) parity proofs within a 50-50 set are 36-19/48-31, 39-21/47-29, and 42-21/50-29 and three examples within a 54-54 set are 37-19/53-35, 41-21/53-33 and 44-23/52-31. We hope to make several examples of such proofs available on the interactive website \cite{WebsitePKA2010}.
\begin{table}[htp]
\vskip-7pt
\begin{center}
\small\addtolength{\tabcolsep}{-3pt}
\centering 
\begin{tabular}{|c|c||cc|cc|cc|cc|cc|cc|cc|cc|cc|cc|} 
\hline 
&&A&&A&&A&&A&&B&&B&&B&&C&&C&&D&\\
\hline
&&B&&C&&D&&E&&C&&D&&E&&D&&E&&E&\\
\hline\hline
A\'&B\'&3&16&2&13&1&14&4&15&25&28&35&36&31&29&32&30&33&34&27&26\\[-3pt]
&&8&17&7&18&5&20&6&19&45&47&42&43&38&37&39&40&41&44&46&48\\[-3pt]
&&10&24&12&22&9&23&11&21&51&50&58&57&56&54&55&53&60&59&49&52\\
\hline
A\'&C\'&2&25&3&28&4&27&1&26&24&22&17&20&16&15&13&14&18&19&23&21\\[-3pt]
&&6&31&5&32&7&30&8&29&42&41&38&40&45&48&47&46&39&37&43&44\\[-3pt]
&&9&35&11&33&10&36&12&34&56&55&51&49&58&59&60&57&50&52&53&54\\
\hline
A\'&D\'&1&38&4&39&3&40&2&37&17&18&16&14&24&21&22&23&13&15&20&19\\[-3pt]
&&7&42&8&41&6&43&5&44&31&32&25&27&35&34&33&36&28&26&30&29\\[-3pt]
&&11&45&9&47&12&46&10&48&58&60&56&53&51&52&50&49&55&54&57&59\\
\hline
A\'&E\'&4&51&1&50&2&49&3&52&16&13&24&23&17&19&18&20&22&21&14&15\\[-3pt]
&&5&56&6&55&8&53&7&54&35&33&31&30&25&26&28&27&32&29&36&34\\[-3pt]
&&12&58&10&60&11&57&9&59&38&39&45&46&42&44&41&43&47&48&40&37\\
\hline
B\'&C\'&15&28&14&25&13&26&16&27&3&2&10&9&8&6&7&5&12&11&1&4\\[-3pt]
&&20&29&19&30&17&32&18&31&37&40&47&48&43&41&44&42&45&46&39&38\\[-3pt]
&&22&36&24&34&21&35&23&33&57&59&54&55&50&49&51&52&53&56&58&60\\
\hline
B\'&D\'&14&37&15&40&16&39&13&38&8&7&3&1&10&11&12&9&2&4&5&6\\[-3pt]
&&18&43&17&44&19&42&20&41&36&34&29&32&28&27&25&26&30&31&35&33\\[-3pt]
&&21&47&23&45&22&48&24&46&54&53&50&52&57&60&59&58&51&49&55&56\\
\hline
B\'&E\'&13&50&16&51&15&52&14&49&10&12&8&5&3&4&2&1&7&6&9&11\\[-3pt]
&&19&54&20&53&18&55&17&56&29&30&28&26&36&33&34&35&25&27&32&31\\[-3pt]
&&23&57&21&59&24&58&22&60&43&44&37&39&47&46&45&48&40&38&42&41\\
\hline
C\'&D\'&27&40&26&37&25&38&28&39&9&11&6&7&2&1&3&4&5&8&10&12\\[-3pt]
&&32&41&31&42&29&44&30&43&15&14&22&21&20&18&19&17&24&23&13&16\\[-3pt]
&&34&48&36&46&33&47&35&45&49&52&59&60&55&53&56&54&57&58&51&50\\
\hline
C\'&E\'&26&49&27&52&28&51&25&50&6&5&2&4&9&12&11&10&3&1&7&8\\[-3pt]
&&30&55&29&56&31&54&32&53&20&19&15&13&22&23&24&21&14&16&17&18\\[-3pt]
&&33&59&35&57&34&60&36&58&48&46&41&44&40&39&37&38&42&43&47&45\\
\hline
D\'&E\'&39&52&38&49&37&50&40&51&1&4&11&12&7&5&8&6&9&10&3&2\\[-3pt]
&&44&53&43&54&41&56&42&55&21&23&18&19&14&13&15&16&17&20&22&24\\[-3pt]
&&46&60&48&58&45&59&47&57&27&26&34&33&32&30&31&29&36&35&25&28\\
\hline 
\end{tabular}
\end{center}\vskip-5pt
\caption{The 100 dual line pairs (DLPs) of the 600-cell.
The DLPs in any row or column provide a tiling of the 600-cell. The letters at
the beginning of each row and column help identify the pair of
24-cells in which each line of a DLP originates. For example, the
line (1 5 9) originates in AA' and DB', while (14 20 23) originates
in AB' and DA'.}
\label{DLP}
\vskip-3pt
\end{table}

\begin{table}[htp]
\vskip-3pt
\addtolength{\tabcolsep}{3pt}
\centering 
\begin{tabular}{|cc|cc|cc|cc|cc|cc|} 
\hline
1&4&2&3&5&7&6&8&9&10&11&12\\[-3pt]
15&14&16&13&18&20&17&19&24&23&22&21\\[-3pt]
56&55&54&53&59&58&57&60&50&52&51&49\\[-3pt]
47&45&46&48&38&37&39&40&44&41&43&42\\[-3pt]
30&29&32&31&36&33&34&35&27&25&26&28\\[-1pt]
\hline
2&3&1&4&6&8&5&7&11&12&9&10\\[-3pt]
43&44&41&42&47&46&45&48&40&38&37&39\\[-3pt]
33&35&36&34&26&25&27&28&29&32&30&31\\[-3pt]
17&18&19&20&24&21&22&23&13&15&16&14\\[-3pt]
52&49&51&50&53&55&54&56&58&57&60&59\\[-1pt]
\hline
5&6&7&8&9&12&10&11&1&3&2&4\\[-3pt]
21&23&24&22&13&14&15&16&17&20&19&18\\[-3pt]
31&32&29&30&34&35&33&36&28&26&27&25\\[-3pt]
50&51&49&52&56&54&53&55&59&60&58&57\\[-3pt]
40&37&39&38&43&41&42&44&46&45&47&48\\[-1pt]
\hline
7&8&5&6&10&11&9&12&2&4&1&3\\[-3pt]
26&27&25&28&32&30&29&31&36&35&33&34\\[-3pt]
16&13&15&14&19&17&18&20&21&22&24&23\\[-3pt]
41&42&43&44&45&48&46&47&39&37&40&38\\[-3pt]
57&59&60&58&49&50&51&52&56&53&54&55\\[-1pt]
\hline
9&11&10&12&1&2&3&4&5&8&6&7\\[-3pt]
19&20&18&17&22&23&21&24&16&14&13&15\\[-3pt]
38&39&40&37&44&42&41&43&47&48&45&46\\[-3pt]
28&25&26&27&31&29&30&32&34&33&36&35\\[-3pt]
53&54&56&55&57&60&58&59&49&51&52&50\\[-1pt]
\hline
10&12&9&11&3&4&1&2&6&7&5&8\\[-3pt]
34&36&33&35&27&28&25&26&30&31&29&32\\[-3pt]
58&60&57&59&51&52&49&50&54&55&53&56\\[-3pt]
22&24&21&23&15&16&13&14&18&19&17&20\\[-3pt]
46&48&45&47&39&40&37&38&42&43&41&44\\[-1pt]
\hline
\end{tabular}
\vskip-1pt
\caption{The 36 Dual-Pentagon-Pairs (DPPs) of the
600-Cell. The DPPs in any row or column provide
a tiling of the 600-cell.}
\label{DPP} 
\end{table}

\clearpage

\subsection*{Acknowledgment}
One of us (M. P.) would like to thank his host Hossein Sadeghpour
at ITAMP. M. P.'s stay at ITAMP was Supported by the US National Science
Foundation through a grant for the Institute for Theoretical Atomic, Molecular,
and Optical Physics (ITAMP) at Harvard University and Smithsonian
Astrophysical Observatory and Ministry of
Science, Education, and Sport of Croatia through the project
No.\ 082-0982562-3160. M. P. carried out his part of the computation on the
cluster Isabella of the University Computing Centre of the University of Zagreb
and on the Croatian National Grid Infrastructure.


\begin{thebibliography}{10}
\bibitem{note1}
We explain some aspects of our terminology for readers unfamiliar with the BKS
  theorem. By a ``ray'' we mean an equivalence class of quantum states that
  differ from each other only by an overall phase. Only orthogonalities between
  states play a role in the BKS theorem, and since orthogonalities are
  unaffected by a change of phase, it is rays rather than states that are of
  relevance for the BKS theorem. The bases are related to the projective
  measurements that one can carry out on the system. The projectors on to the
  rays in a basis form a set of commuting operators, and a joint measurement of
  these projectors causes the system to collapse into the ray associated with
  one of them, with the eigenvalue 1 being returned for this projector and 0
  for the others. This explains why a hidden variables theory attempting to
  simulate quantum mechanics is required to assign a 1 to one projector and 0's
  to the others. The rays and bases in our theoretical discussion correspond to
  states and compatible sets of projective measurements in an actual
  experiment.
\end{thebibliography}
\end{document}